\documentclass[conference,final]{IEEEtran}
\usepackage{amsmath,amssymb}
\usepackage{algorithm,algorithmic}
\usepackage{graphicx}
\newcommand{\LB}{\left(}
\newcommand{\RB}{\right)}
\newcommand{\LSB}{\left[}
\newcommand{\RSB}{\right]}
\newcommand{\htp}{^{\sf H}}
\newcommand{\tp}{^{\sf T}}

\newfont{\bbb}{msbm10 scaled 500}

\newfont{\bb}{msbm10 scaled 1100}
\newcommand{\CC}{\mbox{\bb C}}

\newcommand{\hv}{{\bf h}}

\newcommand{\nv}{{\bf n}}

\newcommand{\xv}{{\bf x}}
\newcommand{\yv}{{\bf y}}

\newcommand{\zerov}{{\bf 0}}

\newcommand{\Gm}{{\bf G}}
\newcommand{\Hm}{{\bf H}}
\newcommand{\Id}{{\bf I}}

\newcommand{\Pm}{{\bf P}}
\newcommand{\Qm}{{\bf Q}}
\newcommand{\Rm}{{\bf R}}
\newcommand{\Sm}{{\bf S}}
\newcommand{\Tm}{{\bf T}}
\newcommand{\Um}{{\bf U}}
\newcommand{\Wm}{{\bf W}}

\newcommand{\Zm}{{\bf Z}}

\newcommand{\Cc}{{\cal C}}

\newcommand{\Nc}{{\cal N}}

\newcommand{\diag}{{\hbox{diag}}}

\renewcommand{\det}{{\hbox{det}}}
\newcommand{\trace}{{\hbox{tr}\,}}

\renewcommand{\arg}{{\hbox{arg}}}

\newcommand{\defines}{{\,\,\stackrel{\scriptscriptstyle \bigtriangleup}{=}\,\,}}

\newtheorem{theorem}{Theorem}
\newtheorem{cor}{Corollary}

\newtheorem{remark}{Remark}[section] 

\newtheorem{assumption}{\indent \bf A}

\begin{document}
\title{Asymptotic Analysis of Double-Scattering Channels}
\author{
\IEEEauthorblockN{
Jakob Hoydis\IEEEauthorrefmark{1}\IEEEauthorrefmark{2},
Romain Couillet\IEEEauthorrefmark{3},
and M\'{e}rouane Debbah\IEEEauthorrefmark{2}} 
\IEEEauthorblockA{\IEEEauthorrefmark{1}Department of Telecommunications,\IEEEauthorrefmark{2}Alcatel-Lucent Chair on Flexible Radio, Sup\'{e}lec, France}
\IEEEauthorblockA{\IEEEauthorrefmark{3}EDF Chair on System Sciences and the Energy Challenge, Centrale Paris-Sup\'elec, France\\
\{jakob.hoydis, romain.couillet, merouane.debbah\}@supelec.fr}
}
\maketitle

\begin{abstract}
We consider a multiple-input multiple-output (MIMO) multiple access channel (MAC), where the channel between each transmitter and the receiver is modeled by the doubly-scattering channel model. Based on novel techniques from random matrix theory, we derive deterministic approximations of the mutual information, the signal-to-noise-plus-interference-ratio (SINR) at the output of the minimum-mean-square-error (MMSE) detector and the sum-rate with MMSE detection, which are almost surely tight in the large system limit. Moreover, we derive the asymptotically optimal transmit covariance matrices. Our simulation results show that the asymptotic analysis provides very close approximations for realistic system dimensions.
\end{abstract}

\section{Introduction}
Most works on wireless multiple-input multiple-output (MIMO) systems share the underlying assumption of a rich scattering environment and, thus, Rayleigh or Rician fading channel matrices with full rank. However, several measurements of outdoor MIMO channels have shown that this assumption fails to hold in certain scenarios, where low-rank channels are observed despite low antenna correlation at the transmitter and receiver (see e.g. \cite{muller01,muller02}). Motivated by these observations, a generalized fading MIMO channel model, the so-called ``doubly-scattering model'' \cite{gesbert02}, was proposed and has since then attracted significant research interest.
A special case of the doubly-scattering model is the keyhole channel \cite{chizik02,almers06} which exhibits null correlation between the entries of the channel matrix but only a single degree of freedom. The existence of such channels in reality was confirmed by measurements in \cite{almers06}.

Several theoretical works have studied the doubly-scattering model so far. The authors of \cite{shin03} derive capacity upper-bounds for the general model and a closed-form expression for the keyhole channel. An asymptotic study of the outage capacity of the multi-keyhole channel was presented in \cite{levin06}.
The diversity order of the doubly-scattering model was considered in \cite{shin08} and it was shown that a MIMO system with $t$ transmit antennas, $r$ receive antennas and $s$ scatterers achieves the diversity of order $trs/\max(t,r,s)$. A closed-from expression of the diversity-multiplexing trade-off (DMT) was derived in \cite{yang11}. Beamforming along the strongest eigenmode over Rayleigh product MIMO channels, i.e., the doubly-scattering model without any form of correlation, was considered in \cite{jin08}. Here, the authors derive exact expressions of the cumulative distribution function (cdf) and the probability density function (pdf) of the largest eigenvalue of the Gramian of the channel matrix and compute closed-form results for the ergodic capacity, outage probability and signal-to-noise-plus-interference-ratio (SINR) distribution. In a later paper \cite{li10}, the MIMO multiple access channel (MAC) with doubly-scattering fading is analyzed. The authors obtain closed-form upper-bounds on the sum-capacity and prove that the transmitters should send their signals along the eigenvectors of the transmit correlation matrices in order to maximize capacity. 

Despite the significant interest in the doubly-scattering channel model, little work has been done to study its asymptotic performance when the channel dimensions grow large. We are only aware of \cite{muller02}, in which a model without transmit and receive correlation is studied relying on tools from free probability theory. Implicit expressions of the asymptotic mutual information and the SINR of the minimum-mean-square-error (MMSE) detector are found. 

In this paper, we consider a MIMO MAC with double-scattering fading in its most general form and derive deterministic approximations of the (ergodic) mutual information, the (ergodic) sum-rate with MMSE detection and the SINR at the output of the MMSE detector. The approximations become almost surely exact as the dimensions of all channel matrices grow large and can be easily numerically computed with negligible computing complexity. In addition, we provide the asymptotically capacity maximizing transmit covariance matrices and present an iterative water-filling algorithm for their computation. Our numerical results suggest that the asymptotic approximations are already very tight for channel dimensions with as little as four transmit and receive antennas and are therefore of clear practical value.

The key idea behind the proofs in this paper is that the doubly-scattering channel model can be interpreted as a Kronecker channel \cite{couillet10} with a \emph{random} receive correlation matrix, which itself is modeled by the Kronecker model. This observation allows us to build upon \cite{couillet10} which provides an asymptotic analysis of the Kronecker channel model with deterministic correlation matrices. We then extend this work by allowing the correlation matrices to be random.
 The results in this paper are obtained through advanced tools from random matrix theory (inspired by \cite{hoydis11,couillet10b}, see also the textbook \cite{couilletRMT} for a comprehensive introduction and a contemporary overview of recent research results) and are hence not only a novel contribution to the field of wireless communications but also to the field of large random matrix theory. We also believe that the developed techniques can be successfully applied to the study of even more involved channel models, such as channels with line-of-sight (LOS) components or MIMO product channels with an arbitrary number of matrices.

\section{System model}
Consider a discrete-time MIMO channel from $K$ transmitters, equipped with $n_k$ ($k=1,\dots,K$) antennas, respectively, to a receiver with $N$ antennas. The channel output vector $\yv\in\CC^N$ at a given time reads
\begin{align}\label{eq:channel}
 \yv &= \sum_{k=1}^K \Hm_{k}\xv_k +\nv
\end{align}
where $\Hm_k\in\CC^{N\times n_k}$ and $\xv_k=[x_{k,1},\dots,x_{k,n_k}]\tp\sim\Cc\Nc(\zerov,\Qm_{k})$, $\Qm_k\in\CC^{n_k\times n_k}$,  are the channel matrix and the transmit vector associated with the $k$th transmitter, and $\nv\sim\Cc\Nc(0,\rho\Id_N)$ is a noise vector. The channel matrices $\Hm_k$ are modeled by the double-scattering model \cite{gesbert02}
\begin{align}\label{eqn:model}
 \Hm_k = \frac{1}{\sqrt{N_k n_k}}\Rm_k^{\frac12}\Wm_{1,k}\Sm_k^{\frac12}\Wm_{2,k}\Tm_k^{\frac12}
\end{align}
where $\Rm_k\in\CC^{N\times N}$, $\Sm_k\in\CC^{N_k\times N_k}$ and $\Tm_k\in\CC^{n_k \times n_k}$ are deterministic correlation matrices, while $\Wm_{1,k}\in\CC^{N\times N_k}$ and $\Wm_{2,k}\in\CC^{N_k\times n_k}$ are independent standard complex Gaussian matrices. Since the distributions of $\Wm_{1,k}$ and $\Wm_{2,k}$ are unitarily invariant we can assume without loss of generality $\Sm_k=\diag(s_{k,1},\dots,s_{k,N_k})$ to be diagonal matrices.
Denote $I_N(\rho)$ the instantaneous normalized mutual information of the channel \eqref{eq:channel} in nats/s/Hz, defined as
\begin{align}
 I_N(\rho) = \frac1N\log\det\LB\Id_N+\frac1\rho\sum_{k=1}^K\Hm_k\Qm_k\Hm_k\htp\RB.
\end{align}
 
\section{Main results}
The notation $N\to\infty$ denotes in the sequel that $N$ and all $N_k$, $n_k$ grow infinitely large, satisfying $ 0 < \lim\inf\frac{N_k}{N} \le \lim\sup \frac{N_k}{N} < \infty$,  $ 0 < \lim\inf\frac{n_k}{N} \le \lim\sup \frac{n_k}{N} < \infty$. 
 These conditions ensure that all matrix dimensions grow at a similar speed. Additionally we need the following technical assumptions:

\vspace{2.5pt}\begin{assumption} For all $k$, $\limsup\lVert\Rm_k\rVert<\infty$, $\limsup\lVert\Sm_k\rVert<\infty$ and $\limsup\lVert\Tm_k\Qm_k\rVert<\infty$, where $\lVert \cdot\rVert$ is the spectral norm.
\end{assumption}\vspace{2.5pt}

Our first theorem introduces a set of $3K$ implicit equations which uniquely determines the quantites $(g_k,\bar{g}_k,\delta_k)$  ($1\le k\le K$). These quantities will be used in the sequel to provide deterministic approximations of $I_N(\rho)$ which become almost surely arbitrarily tight as $N\to \infty$.

\vspace{5pt}\begin{theorem}[Fundamental equations]\label{th:fundequ}
The following system of $3K$ implicit equations in $\bar{g}_k$, $g_k$ and $\delta_k$ ($1\le k \le K$):
\begin{align}\nonumber
 \bar{g}_k &= \frac{1}{n_k}\trace\Tm_k^{\frac12}\Qm_k\Tm_k^{\frac12}\LB g_k\Tm_k^{\frac12}\Qm_k\Tm_k^{\frac12}+\Id_{n_k}\RB^{-1}\\\label{eq:fundequ}
 g_k &= \frac{1}{n_k}\sum_{j=1}^{N_k}\frac{s_{k,j}\delta_k}{1+\bar{g}_k s_{k,j}\delta_k}\\\nonumber
 \delta_k &= \frac{1}{N_k}\trace\Rm_k\LB\sum_{k=1}^K \frac{n_k}{N_k}\frac{\bar{g}_k g_k}{\delta_k}\Rm_k+\rho\Id_N\RB^{-1}
\end{align}
has a unique solution satisfying $\bar{g}_k,g_k,\delta_k>0$ for all $k$ and $\rho>0$.
\end{theorem}\vspace{5pt}

\begin{remark}
 One can also prove that $\bar{g}_k$, $g_k$ and $\delta_k$ can be computed by a classical fixed-point algorithm which iteratively computes \eqref{eq:fundequ}, starting from some arbitrary initialization $\bar{g}_k,g_k,\delta_k>0$. This algorithm generally converges in a few iterations (depending on the system size) and does not pose any computational challenge.
\end{remark}\vspace{10pt}

The next theorem provides a deterministic, asymptotically tight approximation of the (ergodic) mutual information based on the quantites $(g_k,\bar{g}_k,\delta_k)$ as provided by Theorem~\ref{th:fundequ}.

\vspace{10pt}\begin{theorem}[Mutual information]\label{th:mutinf}
\begin{align*}
 (i)\quad &I_N(\rho)-\bar{I}_N(\rho) \xrightarrow[N\to\infty]{\text{a.s.}} 0\\
 (ii)\quad &\mathbb{E}I_N(\rho)-\bar{I}_N(\rho) \xrightarrow[N\to\infty]{} 0
\end{align*}
where
\begin{align*}
  \bar{I}_N(\rho) \ = &\ \frac1N\log\det\LB\Id_N + \frac{1}{\rho}\sum_{k=1}^K\frac{n_k}{N_k}\frac{\bar{g}_k g_k}{\delta_k}\Rm_k\RB\\
 &\ + \frac1N\sum_{k=1}^K\LSB \log\det\LB\Id_{N_k}+\bar{g}_k\delta_k\Sm_k\RB\right. \\
 &\ +\left. \log\det\LB\Id_{n_k} + g_k\Tm_k^{\frac12}\Qm_k\Tm_k^{\frac12}\RB 
 -2n_k g_k\bar{g}_k\RSB
\end{align*}
and $g_k,\bar{g}_k,\delta_k$ are the unique positive solutions to \eqref{eq:fundequ}.
\end{theorem}\vspace{10pt}

The following result allows us to compute the asymptotically optimal precoding matrices $\Qm_k$ which maximize $\bar{I}_N(\rho)$ under individual transmit power constraints.

\vspace{10pt}\begin{theorem}[Optimal power allocation]\label{th:optpow}
The solution to the following optimization problem:
\begin{align*}
 \LB\bar{\Qm}_1^*,\dots,\bar{\Qm}_K^*\RB\ =\ & \arg\max_{\Qm_1,\dots,\Qm_k} \bar{I}_N(\rho)\\
 &\ \text{s.t.}\quad \frac{1}{n_k}\trace \Qm_k\le P_k\ \forall k
\end{align*}
is given as $\bar{\Qm}_k^* = \Um_k\bar{\Pm}_k^* \Um_k\htp$, where $\Um_k\in\CC^{n_k\times n_k}$ is defined by the spectral decomposition of $\Tm_k=\Um_k\diag(t_{k,1},\dots,t_{k,n_k})\Um_k\htp$ and $\bar{\Pm}_k^*=\diag(\bar{p}^*_{k,1},\dots,\bar{p}^*_{k,n_k})$ is given by the water-filling solution:
\begin{align}\label{eq:wf}
 \bar{p}^*_{k,j} = \LB\mu_k - \frac{1}{g_k^*t_{k,j}}\RB^+
\end{align}
where $\mu_k$ is chosen to satisfy $\frac1{n_k}\trace\bar{\Pm}_k^*=P_k$ and $g_k^*$ is given by Theorem~\ref{th:fundequ} for $\Qm_k=\bar{\Qm}^*_k$.
\end{theorem}\vspace{10pt}

\begin{remark}
 The optimal power allocation matrices $\bar{\Pm}_k^*$  can be calculated by the iterative water-filling Algorithm~1 (see \cite[Remark 2]{couillet10} and \cite[Remark 5]{hoydis11} for a discussion of the convergence of this algorithm).
\begin{algorithm}
\caption{Iterative water-filling algorithm}
\label{alg:wf}
 \begin{algorithmic}[1]
 \STATE Let $\epsilon>0$, $n=0$ and $\bar{p}^{*,0}_{k,j}=P_k$ for all $k,j$.
 \REPEAT
 \STATE For all $k$, compute $g_k^{*,n}$ according to Theorem~\ref{th:fundequ} with matrices $\Qm_k=\Um_k\diag\LB \bar{p}_{k,j}^{*,n}\RB\Um_k\htp$.
 \STATE For all $k,j$, calculate  $\bar{p}_{k,j}^{*,n+1}=\LB\mu_k - \frac{1}{g_k^{*,n}t_{k,j}}\RB^{+}$, with $\mu_k$ such that $\frac1{n_k}\sum_{j=1}^{n_k}\bar{p}_{k,j}^{*,n+1}=P_k$.
 \STATE $n=n+1$
 \UNTIL{$\max_{k,j} |\bar{p}_{k,j}^{*,n}-\bar{p}_{k,j}^{*,n-1}|\le\epsilon$} 
\end{algorithmic}
\end{algorithm}
\end{remark}\vspace{10pt}

The last two results of this correspondence provide deterministic approximations of the SINR at the output of the MMSE detector and the sum-rate with MMSE detection.

\vspace{10pt}\begin{theorem}[SINR of the MMSE detector]\label{th:sinr}
Assume $\Qm_k=\Id_{n_k}$ and $\Tm_k=\diag(t_{k,1},\dots,t_{k,n_k})$ for all $k$ and let $\gamma_{k,j}$ be the SINR at the output of the MMSE detector related to the transmit symbol $x_{k,j}$, given by
\begin{align*}
 \gamma_{k,j} = \hv_{k,j}\htp\LB\sum_{i=1}^K\Hm_i\Hm_i\htp - \hv_{k,j}\hv_{k,j}\htp + \rho\Id_N\RB^{-1}\hv_{k,j}.
\end{align*}
Then
\begin{align*}
 \gamma_{k,j}-\bar{\gamma}_{k,j}\xrightarrow[N\to\infty]{\text{a.s.}} 0
\end{align*}
where $\bar{\gamma}_{k,j}= t_{k,j}g_k $ and $g_k$ is by given by Theorem~\ref{th:fundequ}.
\end{theorem}

\vspace{10pt}\begin{remark}
 The theorem is also valid under the more general assumptions $\Tm_k=\Um_k\diag(t_{k,1},\dots,t_{k,n_k})\Um_k\htp$ and $\Qm_k=\Um_k\diag(p_{k,1},\dots,p_{k,n_k})\Um_k\htp$. We can then simply define the matrices $\Tm_k'=\diag(t_{k,1}p_{k,1},\dots,t_{k,n_k}p_{k,n_k})$ and $\Qm_k'=\Id_{N_k}$ for which the theorem holds.
\end{remark}

\vspace{10pt}\begin{cor}[Sum-rate with MMSE decoding]\label{cor:rate}
Under the same assumptions as in Theorem~\ref{th:sinr}, let 
\begin{align*}
 R(\rho) = \frac1N\sum_{k=1}^K\sum_{j=1}^{n_k}\log(1+\gamma_{k,j}).
\end{align*}
Then,
\begin{align*}
 (i)&\quad R(\rho) - \frac1N\sum_{k=1}^K\sum_{j=1}^{n_k}\log(1+t_{k,j}g_k)\xrightarrow[N\to\infty]{\text{a.s.}} 0\\
(ii)&\quad \mathbb{E}R(\rho) - \frac1N\sum_{k=1}^K\sum_{j=1}^{n_k}\log(1+t_{k,j}g_k)\xrightarrow[N\to\infty]{\text{a.s.}} 0
\end{align*}
where $g_k$ is given by Theorem~\ref{th:fundequ}.
\end{cor}\vspace{10pt}

\subsection{The Rayleigh product channel}
A special case of the double-scattering channel is the Rayleigh product MIMO channel \cite{jin08} which does not exhibit any form of correlation between the transmit and receive antennas or the scatterers. For this model, the Theorems~\ref{th:fundequ}, \ref{th:mutinf} and \ref{th:sinr} can be given in closed-from as shown in the next corollary.

\begin{cor}[Rayleigh product channel]\label{cor:rayprod}
 For all $k$, let $N_k=S$, $n_k = N$ and assume $\Tm_k=\Id_N$, $\Sm_k=\Id_S$ and $\Rm_k=\Id_N$. Then, 
\begin{align*}
 \bar{I}_N(\rho)\ =&\ \log\LB1+\frac1\rho\frac{NK}{S}\bar{g}\LB\bar{g}+\frac{S}{N}- 1\RB\RB\\
&\quad- \ \frac{KS}{N}\log\LB1+\frac NS\LB\bar{g}-1\RB\RB\\
&\quad-K\log\LB\bar{g}\RB -2K\LB1-\bar{g}\RB
\end{align*}
and 
\begin{align*}
 \bar{\gamma}_{k,j} = \frac{1-\bar{g}}{\bar{g}}
\end{align*}
where $\bar{g}$ is the unique solution to
\begin{align}\label{eq:fundequsimple}\nonumber
 &\bar{g}^3 - \bar{g}^2\LB2-\frac SN - \frac1K\RB \\&\quad+\bar{g}\LB1-\frac SN - \frac1K + \frac S{NK}(1+\rho)\RB - \frac{S}{NK}\rho\ =\ 0
\end{align}
such that  $\delta \defines (1-\bar{g})/(\bar{g}(\bar{g}+S/N-1))>0$ and $g \defines(1-\bar{g})/\bar{g}>0$. 
\end{cor}\vspace{10pt}

Note that similar expressions for the asymptotic mutual information and MMSE-SINR have been obtained in \cite{muller02} by means of free probability theory. However, their results require the numerical solution of a third order differential equation.

\section{Numerical examples}
\begin{figure}
\centering
\includegraphics[width=0.47\textwidth]{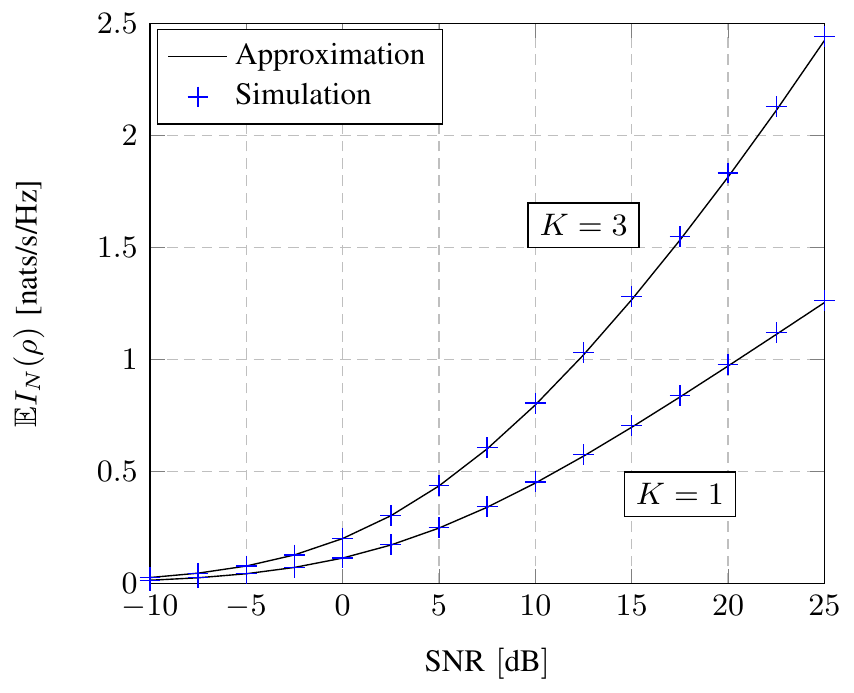}
\caption{Ergodic mutual information $\mathbb{E}I_N(\rho)$ of the multi-keyhole channel and its asymptotic approximation $\bar{I}_N(\rho)$ versus SNR.\label{fig:keyhole}}
\end{figure}

As a first example, we consider the ``multi-keyhole channel'', i.e., $N_k=1$, $\Sm_k=1$, $\Rm_k=\Id_N$, $\Tm_k=\Qm_k=\Id_{n_k}$,
for $N=n_k=4$. The signal-to-noise-ratio (SNR) is denoted by $\text{SNR}=1/\rho$.
Fig.~\ref{fig:keyhole} depicts the normalized ergodic mutual information $\mathbb{E}I_N(\rho)$ and its asymptotic approximation $\bar{I}_N(\rho)$ versus SNR for a single ($K=1$) and multiple ($K=3$) ``keyholes''. Surprisingly, the match between both results is almost perfect although the channel dimensions are very small.

\begin{figure}
\centering
\includegraphics[width=0.47\textwidth]{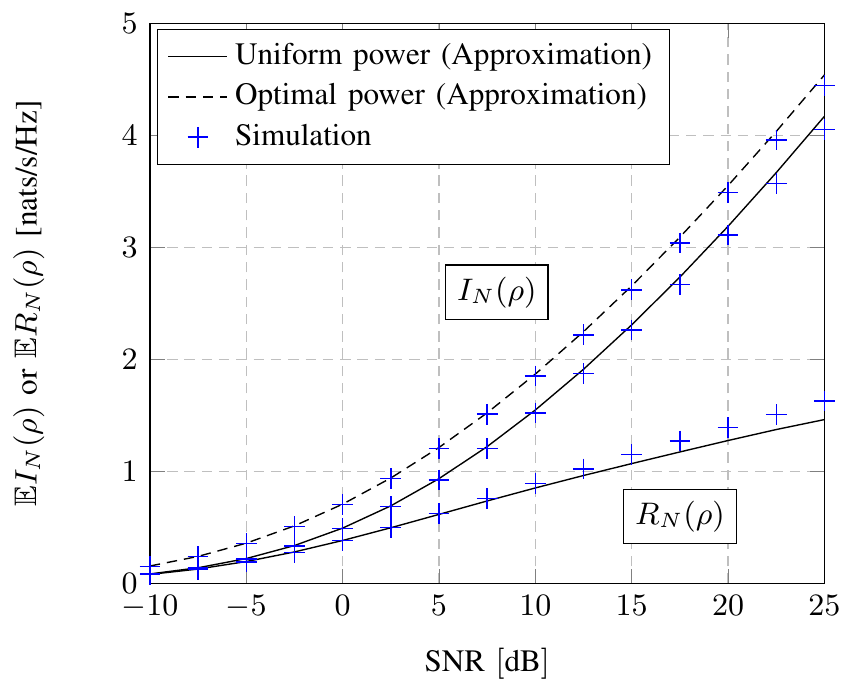}
\caption{Ergodic mutual information $\mathbb{E}I_N(\rho)$ and sum-rate $\mathbb{E}R_N(\rho)$ of the multiple access channel and their asymptotic approximations $\bar{I}_N(\rho)$ and $\bar{R}_N(\rho)$ versus SNR.\label{fig:mac}}
\end{figure}

As a second example, we consider a multiple access channel from $K=3$ transmitters, assuming the double-scattering model in \cite{gesbert02}. Under this model, the correlation matrices are given as  $\Rm_k = \Gm(\phi_{r,k},d_{r,k},N_k)$, $\Sm_k=\Gm(\phi_{s,k},d_{s,k},N_k)$ and $\Tm_k=\Gm(\phi_{t,k},d_{t,k},N_k)$, where $\Gm(\phi,d,n)$ is defined as
\begin{align*}
 \LSB\Gm(\phi,d,n)\RSB_{k,l} = \frac1n \sum_{j=\frac{1-n}{2}}^{\frac{n-1}{2}}\exp\LB{\bf{i}}2\pi d(k-l)\sin\LB\frac{j\phi}{1-n}\RB\RB.
\end{align*}
The values $\phi_{t,k}$ and  $\phi_{r,k}$ determine the angular spread of the radiated and received signals, $d_{t,k}$ and $d_{r,k}$ are the antenna spacings at the $k$th transmitter and receiver in multiples of the signal wavelength, and $N_k$ can be seen as the number and $d_{s,k}$ as the spacing of the scatterers. For simplicity, we assume $N=4$, $P_k=1/n_k$, $N_k=11$, $n_k=3$, $d_{t,k}=d_{r,k} = 0.25$ and $d_{s,k}=50$ for all $k$. We further assume $\phi_{r,k}=\phi_{t,k}$ for all $k$, with $\phi_{r,k}\in\{\pi/4,\pi/2,\pi\}$ and $\phi_{s,k}=\pi/8$. Fig.~\ref{fig:mac} shows $\mathbb{E}I_N(\rho)$ and $\bar{I}_N(\rho)$ with uniform and optimal power allocation versus SNR. Again, our asymptotic results yield very tight approximations for even small system dimensions. For comparison, we also provide the sum-rate with MMSE detection $\mathbb{E}R_N(\rho)$ and its deterministic approximation $\bar{R}_N$. We observe a good fit between both results at low SNR values, but a slight mismatch for higher values. This is due to a slower convergence of the SINR $\gamma_{k,j}$ to its deterministic approximation $\bar{\gamma}_{k,j}$.

\section{Conclusion}
We have studied a MIMO MAC with doubly-scattering fading channels. Under the assumption that the dimensions of all channel matrices grow infinitely large, we have derived almost surely tight deterministic approximations of the mutual information, the SINR of the MMSE detector and the sum-rate with MMSE detection.
In addition, we have provided an iterative water-filling algorithm to compute the asymptotically optimal transmit covariance matrices. Our numerical results show that the asymptotic analysis provides very close approximations for very small system dimensions with as little as four transmit and received antennas. We believe that the techniques used in this paper could be succesfully applied to the study of even more involved channel models.

\section{Appendix}
\begin{IEEEproof}[Sketch of proof of Theorem \ref{th:fundequ}]
The key idea is that the doubly scattering model can be considered as the Kronecker channel model \cite{couillet10} with random correlation matrices. For the Kronecker model, the matrices $\Hm_k$ are given as
\begin{align}\label{eq:kron}
 \Hm_k = \frac1{\sqrt{n_k}}\Zm_k\Wm_{2,k}\Tm_k^{\frac12}
\end{align}
where $\Zm_k\in\CC^{N\times N_k}$ is a deterministic matrix and $\Wm_{2,k}$ and $\Tm_k$ are as defined in \eqref{eqn:model}. The fundamental equations (cf. Theorem~\ref{th:fundequ}) for this model are given by \cite[Corollary 1]{couillet10}
\begin{align}\nonumber
 \bar{e}_k &= \frac1{n_k}\trace\Tm_k^{\frac12}\Qm_k\Tm_k^{\frac12}\LB e_k\Tm_k^{\frac12}\Qm_k\Tm_k^{\frac12} + \Id_{n_k}\RB^{-1}\\\label{eq:fundeqkron}
e_k &= \frac1{n_k}\trace\Zm_k\Zm_k\htp\LB\sum_{i=1}^K\bar{e}_i\Zm_i\Zm_i\htp + \rho\Id_N\RB^{-1}.
\end{align}
Assume now $\Zm_k$ to be random and modeled as
\begin{align}\label{eq:corkron}
 \Zm_k = \frac1{\sqrt{N_k}} \Rm_k^{\frac12}\Wm_{1,k}\Sm_k^{\frac12}.
\end{align}
Notice first that the expressions of the quantities $\bar{e}_k$ are unaffected by this assumption. Second, $e_k$ have become
random quantities and it is our goal to find deterministic approximations $g_k$ of $e_k$, such that $e_k-g_k\xrightarrow[]{\text{a.s.}} 0$ as $N\to\infty$. Following similar steps as in the proof of \cite[Theorem 4]{hoydis11}, one can now show that 
\begin{align}\nonumber
 \max_k\left|\bar{e}_k -\bar{g}_k \right| &\xrightarrow[N\to\infty]{\text{a.s.}} 0\\\label{eq:convek}
\max_k\left|e_k -g_k \right| &\xrightarrow[N\to\infty]{\text{a.s.}} 0
\end{align}
where $\bar{g}_k$ and $g_k$ satisfy \eqref{eq:fundequ}. The proof of uniqueness of such solutions relies on arguments of so called \emph{standard functions} \cite{yates95} and follows similar steps as in \cite[Proof of Theorem 3]{hoydis11} or \cite[Proof of Theorem 1]{couillet10b}.
\end{IEEEproof}

\vspace{10pt}\begin{IEEEproof}[Sketch of proof of Theorem \ref{th:mutinf}]
 We rely again on the observation that the doubly scattering model can be considered as the Kronecker channel model \cite{couillet10} with random correlation matrices  $\Zm_k$ (cf. \eqref{eq:kron} and \eqref{eq:corkron}). A deterministic equivalent $\bar{I}^\text{Kron}_N(\rho)$ of the mutual information for the Kronecker model \eqref{eq:kron} was provided in \cite[Theorem 2]{couillet10}. Here, $\bar{I}^\text{Kron}_N(\rho)$ is defined as a function of the quantities $\bar{e}_k$ and $e_k$ (given as the unique solutions to \eqref{eq:fundeqkron}) which reads:
\begin{align}\nonumber
 &\bar{I}^\text{Kron}_N(\rho) = \frac1N\log\det\LB\Id_N + \frac1\rho\sum_{k=1}^K\bar{e}_k\Zm_k\Zm_k\htp\RB\\\label{eq:mutinfkron}
&\quad + \sum_{k=1}^K\frac1N\log\det\LB\Id_{n_k}+e_k\Tm_k^{\frac12}\Qm_k\Tm_k^{\frac12}\RB  - \frac1N \sum_{k=1}^K n_ke_k\bar{e}_k.
\end{align}
Due to the almost sure convergence of $e_k-g_k\to 0$ and $\bar{e}_k-\bar{g}_k\to 0$ established in Theorem~\ref{th:fundequ}, we can simply replace $e_k$ and $\bar{e}_k$ by $g_k$ and $\bar{g}_k$, respectively. It remains now to find a deterministic equivalent of the first term $\frac1N\log\det\LB\Id_N + \frac1\rho\sum_{k=1}^K\bar{e}_k\Zm_k\Zm_k\htp\RB$, which is random since the matrices $\Zm_k$ are random. However, this term is nothing but the mutual information of \emph{another} Kronecker channel with matrices $\tilde{\Hm}_k=\sqrt{\frac{\bar{e}_k}{N_k}}\Rm_k^{\frac12}\Wm_{1,k}\Sm_k^{\frac12}$. Hence we can apply again \cite[Theorem 2]{couillet10} to find its deterministic equivalent. Combining both results yields $\bar{I}_N(\rho)$ and terminates the proof of $(i)$. Denote $\Omega$ the probability space engendering the sequences $\{\Wm_{1,1},\dots,\Wm_{1,K},\Wm_{2,1},\dots,\Wm_{2,K}\}$. Then, on a sub-space of $\Omega$ of measure $1$, we have by $(i)$: $I_N(\rho) - \bar{I}_N(\rho) \to 0$ as $N\to\infty$. Integrating this expression over $\Omega$ implies by the dominated convergence theorem \cite{billingsley} part $(ii)$.
\end{IEEEproof}

\vspace{10pt}\begin{IEEEproof}[Sketch of proof of Theorem \ref{th:optpow}]
Similar to the proof of \cite[Proposition 3]{couillet10}, one can show that the covariance matrices $\bar{\Qm}^*_k$ should align to the eigenvectors of the transmit correlation matrices $\Tm_k$ to maximize $\bar{I}_N(\rho)$, i.e.,   $\bar{\Qm}^*_k = \Um_k\bar{\Pm}_k^* \Um_k\htp$. Note that it was also proved in \cite[Theorem 1]{li10} that these signaling directions are optimal to maximize $\mathbb{E}I_N(\rho)$. One can then further show that
\begin{align*}
 \frac{d \bar{I}_N(\rho)}{d \bar{p}^*_{k,j}} = \frac{g_k^*t_{k,j}}{1+g_k^* t_{k,j}\bar{p}^*_{k,j}}\quad \forall k,j
\end{align*}
and $\frac{d^2 \bar{I}_N(\rho)}{d (\bar{p}^*_{k,j})^2}<0 $. Since $\bar{I}_N(\rho)$ is hence strictly concave in $\bar{p}^*_{k,j}$ it follows from the KKT conditions \cite{boydcvx} that $\bar{\Pm}_k^*$ is given by the water-filling solution \eqref{eq:wf} with power constraint $\frac1{n_k}\trace\bar{\Qm}^*_k=\frac1{n_k}\trace\bar{\Pm}^*_k=P_k$.
\end{IEEEproof}

\vspace{10pt}\begin{IEEEproof}[Sketch of proof of Theorem~\ref{th:sinr}]
Assume that $\Hm_k$ are given by the Kronecker model \eqref{eq:kron} with deterministic matrices $\Zm_k$. It follows from standard lemmas of random matrix theory (see e.g. \cite[Appendix C]{hoydis11}, that the following limit holds:
\begin{align*}
 \gamma_{k,j} - t_{k,j}\frac{1}{n_k}\trace\Zm_k\Zm_k\htp\LB\sum_{i=1}^K\Hm_i\Hm_i\htp + \rho\Id_{n_k}\RB\xrightarrow[N\to\infty]{\text{a.s.}}0.
\end{align*}
Applying \cite[Theorem 1]{couillet10} to the second term yields
\begin{align*}
\gamma_{k,j}-t_{k,j}\frac{1}{n_k}\trace\Zm_k\Zm_k\htp\LB\sum_{i=1}^K\bar{e}_i\Zm_i\Zm_i\htp + \rho\Id_{n_k}\RB\xrightarrow[N\to\infty]{\text{a.s.}}0
\end{align*}
where $\bar{e}_i$ are given as the solutions to \eqref{eq:fundeqkron}. Notice from \eqref{eq:fundeqkron} that the second term is equal to $t_{k,j}e_k$.
Assume now the matrices $\Zm_k$ random and given by \eqref{eq:corkron}. Thus, we have by \eqref{eq:convek}
\begin{align*}
 \gamma_{k,j} - t_{k,j}g_k\xrightarrow[N\to\infty]{\text{a.s.}}0.
\end{align*}

\end{IEEEproof}

\vspace{10pt}\begin{IEEEproof}[Sketch of proof of Corollary~\ref{cor:rate}]
Part $(i)$ follows directly from Theorem~\ref{th:sinr} and the continuous mapping theorem \cite[Theorem 2.3]{vdv}.  Part $(ii)$ follows from the same arguments as in the proof of Theorem~\ref{th:optpow} $(ii)$.   
\end{IEEEproof}

\vspace{10pt}\begin{IEEEproof}[Sketch of roof of Corollary~\ref{cor:rayprod}]
One can show by straight-forward but tedious calculations that the fundamental equations  \eqref{eq:fundequ} can be reduced to a single implicit equation \eqref{eq:fundequsimple} if $\Tm_k=\Id_N$, $\Sm_k=\Id_S$ and $\Rm_k=\Id_N$ for all $k$. Replacing $g_k$ and $\delta_k$ by $g$ and $\delta$ in the expressions of $I_N(\rho)$ and $\bar{\gamma}_{k,j}$, respectively, leads to the desired result.
\end{IEEEproof}

\bibliographystyle{IEEEtran}
\bibliography{IEEEabrv,bibliography}
\end{document}